\documentclass[useAMS,usenatbib]{emulateapj}

\def\simless{{\th \rlap{\raise 0.5ex\hbox{$\scriptstyle  {<}$}}
    {\lower 0.3ex\hbox{$\scriptstyle  {\sim}$}} \th }}  
\def\simgreat{{\th \rlap{\raise 0.5ex\hbox{$\scriptstyle  {>}$}}
    {\lower 0.3ex\hbox{$\scriptstyle  {\sim}$}} \th }}  
\def\greateq{{\th \rlap{\raise 0.5ex\hbox{$\scriptstyle  {>}$}}
    {\lower 0.3ex\hbox{$\scriptstyle  {-}$}} \th }}  
\def\lesseq{{\th \rlap{\raise 0.5ex\hbox{$\scriptstyle  {<}$}}
    {\lower 0.3ex\hbox{$\scriptstyle  {-}$}} \th }}  
\def\th{\thinspace}
\def\ts{{\raise 0.3ex\hbox{$\scriptstyle {\th \sim \th }$}}}

\usepackage{epsfig}

\newcommand{\msun}{\,M_{\odot}}

\newcommand{\nudot}{\dot{\nu}}

\newcommand{\Mdot}{\dot{M}}
\newcommand{\avmdot}{\langle\Mdot\rangle}

\newcommand{\source}{IGR J17480--2446 }

\def\hide#1{}

\begin{document}


\shorttitle{Terzan 5}


\title{The Peculiar Evolutionary History of IGR J17480-2446 in Terzan 5}

\author{Alessandro Patruno \altaffilmark{1}, M. Ali Alpar
  \altaffilmark{1,2}, Michiel van der Klis \altaffilmark{1}, Ed P.J. van
  den Heuvel\altaffilmark{1}} \altaffiltext{1}{Astronomical Institute
  ``Anton Pannekoek,'' University of Amsterdam, Science Park 904, 1098
  SJ Amsterdam, The Netherlands;} \altaffiltext{2}{Faculty of Engineering and Natural Sciences, Sabanc{\i} University,  Orhanl{\i}, Tuzla 34956, \.{I}stanbul, Turkey}

\email{a.patruno@uva.nl}

\begin{abstract}

  The low mass X-ray binary (LMXB) \source in the globular cluster Terzan 5
  harbors an 11 Hz accreting pulsar. This is the first object
  discovered in a globular cluster with a pulsar spinning at such low
  rate. The accreting pulsar is anomalous because its characteristics
  are very different from the other five known slow accreting pulsars
  in galactic LMXBs. Many features of the 11 Hz
  pulsar are instead very similar to those of accreting millisecond
  pulsars, spinning at frequencies $>\,100$Hz.  Understanding this
  anomaly is valuable because \source can be the only accreting
  pulsar discovered so far which is in the process of becoming an
  accreting millisecond pulsar. We first verify that the neutron star
  (NS) in \source is indeed spinning up by carefully analyzing X-ray data
  with coherent timing techniques that account for the presence of
  timing noise.  We then study the present Roche lobe overflow epoch
  and the two previous spin-down epochs dominated by magneto dipole
  radiation and stellar wind accretion. We find that \source is very
  likely a mildly recycled pulsar and suggest that it has started a
  spin-up phase in an exceptionally recent time, that has lasted less
  than a few $10^7$ yr. We also find that the total age of the
  binary is surprisingly low ($\simless10^8$ yr) when considering
  typical parameters for the newborn NS and propose
  different scenarios to explain this anomaly.

\end{abstract}


\section{Introduction}\label{intro}

According to the recycling scenario
(\citealt{alp82},~\citealt{rad82}), millisecond radio pulsars are
produced by spin-up of the neutron stars (NSs) via transfer of angular
momentum through accretion in low mass X-ray binaries (LMXBs). During
this phase, channeled accretion of plasma along the NS magnetic field
lines can produce X-ray pulsations that reveal the spin frequency of the
accreting NS.  A confirmation of this scenario came with the discovery
of the first accreting millisecond X-ray pulsar SAX J1808.4-3658,
spinning at a frequency of about 401 Hz \citep{wij98}.  Thirteen more
accreting millisecond pulsars have been discovered so far with spin
frequencies ranging from about 180 up to 600 Hz.

In 2010 October the new accreting pulsar IGR J17480-2446 has been
discovered in the globular cluster Terzan 5 \citep{bor10, str10}.
\citet{poo10} localized the source in outburst with \textit{Chandra}
observations and identified it with a previously known candidate
quiescent LMXB, named CX25 in the X-ray source catalog of
\citet{hei06}. The source is in core of Terzan 5, one of the densest
among globular clusters \citep{fer09}.

The accreting pulsar spins with a frequency of 11 Hz and the LMXB has
a period of 21.3 hr around a companion with mass $M>0.4\,M_{\odot}$
\citep{pap11}. Although this pulsar is not a millisecond one, its
location in a globular cluster makes it a unique system whose
properties are of particular importance to study the recycling
mechanism. \source is different from the other five known slow
accreting pulsars in LMXBs (4U 1626$-$67, 2A 1822$-$371, GRO 1744-28,
GX 1+4 and Her X-1) with properties which are in between those systems
and the accreting millisecond pulsars \citep{lin12}.  Slow accreting
pulsars are a very heterogeneous group of LMXBs, with white dwarf,
main sequence and giant companions and orbital periods ranging from
less than 1 hr to more than 11 days. Despite this, the five slow
pulsars have a high magnetic field $B\sim10^{11}-10^{13}\rm\,G$ and do
not show thermonuclear bursts. This high $B$ field is difficult to
reconcile with the long lifetime of these slow systems, and several
scenarios have been proposed (see, for example, \citealt{ver90} for a
discussion).

\source instead shows thermonuclear bursts (\citealt{che10},
\citealt{lin11}) with burst oscillations phase locked with the
accretion powered pulsations \citep{cav11}, and the $B$ field has been
constrained to be in the range $10^9-10^{10}$G (\citealt{mil11},
\citealt{cav11}, \citealt{pap11}).  All these field estimates point
toward a mildly recycled pulsar, since accreting millisecond pulsars
have shown so far field strengths of the order of $10^8$G (in
agreement with the $B$ fields of radio millisecond pulsars). Phase
locking between burst oscillations and accretion powered pulsations
has also been observed in at least one accreting millisecond pulsar
(\citealt{str03}, \citealt{wat08}).
The 11 Hz rotation rate is, however ,still unusually slow for an old NS,
the typical rotation periods for accreting or rotation powered pulsars
in globular clusters and old stellar populations being in the
millisecond range. 

The reason for old NSs possessing dipole magnetic fields smaller by
about three orders of magnitude from the typical $10^{12} \rm\,G$
fields of young NSs has elicited several explanations involving binary
evolution. The burial of the field under accreted material has been
proposed as a viable mechanism to dissipate the field via ohmic decay
in the heated NS crust (see for example
\citealt{rom90}). This mechanism, however, has a dramatic
consequence for the final magnetic field: as the accreted matter keeps
falling on the NS surface, the magnetic field lines will be pushed to 
deeper (and colder) regions of the crust where the conductivity is
much larger, thus freezing at the crust-core interface. This will
leave a residual field that is compatible with the observed weak
fields of millisecond pulsars.

Another particularly simple and elegant explanation relies on the
superfluid-superconducting nature of the NS interior, requiring that
the rotation is carried by quantized vortex lines in the neutron
superfluid and the magnetic flux by quantized flux lines in the proton
superconductor \citep{sri90}. Spin down of the star is achieved by
outward motion of the vortex lines, which inevitably entangle and drag
flux lines with them. Srinivasan et al.  noted that many NS binary
evolution scenarios have an early first mass transfer epoch when the
companion, not yet filling its Roche lobe, is losing mass by a stellar
wind, some of which is captured by the NS in a weak quasi spherical
accretion with low specific angular momentum. The NS will be spinning
down in this epoch, thus reducing its dipole field in proportion to
its rotation rate. The wind spin-down epoch ends when the companion
fills its Roche lobe. The ensuing disk accretion now starts to spin-up
the NS.

If IGR J17480-2446 is a primordial binary, then its spin frequency is
surprising low (the globular cluster Terzan 5 is about 12 Gyr old, see
\citealt{fer09}) and requires investigation to understand whether this
is the consequence of a peculiar evolutionary history.  At present
\source is likely to be in its spin-up epoch \citep{pap11}. The
results reported by \citet{pap11} are however affected by the usual
problem of X-ray timing noise in the pulse time of arrivals (TOAs) that when
not taken properly into account might strongly affect the significance
of the spin frequency derivatives (see for example the case of the
other three accreting millisecond X-ray pulsars in
\citealt{har08,har09}, \citealt{pat10} \citealt{has11}).  In this work
we will re-investigate the existence of a spin-up phase by taking into
account the effect of timing noise. If the spin-up is confirmed, then
the comparatively long spin period indicates either that \source is at
an exceptionally early phase of the spin-up epoch or that it is close
to spin equilibrium, with an unusually long equilibrium period
reflecting a magnetic field stronger than in most NSs in LMXB.

The aperiodic variability can also help to understand the peculiar
behavior of the source. \citet{alt10} reported several quasi-periodic 
oscillations (QPOs) at
frequencies ranging from about 48 up to 815 Hz.  This sequence of
frequencies make possible several alternative estimations of the inner
disk radius at present. Each such model gives an estimate of the
dipole magnetic field, equilibrium rotation rate and expected spin-up
rate.

The location of \source in the globular cluster Terzan 5 allows also
to place stronger constraints on the evolutionary history of the
binary since only a narrow range of donor masses is allowed. The
distance of the system is also constrained by the globular cluster and
leads to a precise determination of the X-ray luminosity. In
Section~\ref{xray} we describe the X-ray observations carried during
the 2010 outburst, and in Section~\ref{results} we verify the existence
of a spin-up phase.  In Section~\ref{evolution} we present the
evolutionary history of \source. We discuss three distinct epochs: in
the initial two the NS has been spun down first via magneto dipole
radiation and then by wind accretion, whereas in the third epoch the
NS is spun-up via accretion of mass provided by the donor in Roche
lobe overflow (RLOF).  The evolution of the magnetic moment of the NS
through the earlier dipole and wind spin-down epochs is also
discussed.  In Section~\ref{constraints} we place further constraints
on the binary parameters by using the available information on the
globular cluster Terzan 5 and the results of the evolutionary history.
We conclude with a discussion of the results in
Section~\ref{discussion} and infer that \source is in an exceptionally
early RLOF phase that explains why its spin frequency is still so
small when compared to accreting millisecond pulsars.

\section{X-ray Observations}\label{xray}

We reduced all pointed X-ray observations taken with the \textit{Rossi
  X-ray Timing Explorer} (\textit{RXTE}) by the Proportional Camera
Array~\citep{jah06} in event mode. Only photons with energies
between 2.5 and 15.6 keV are retained (absolute channels 5$-$37). The
time resolution of the observations is $2^{-13}$ s (for $122\mu s$
Event data) and $2^{-20}$ s (for Good Xenon data) rebinned to
$2^{-13}$ s to make the data sampling uniform. The photons detected
span an observing window between 2010 October 13 and November 19.
The total duration of the outburst is longer than the duration of the
\textit{RXTE} observations, and it has been constrained to be 
$\approx 55$ days \citep{cav11} or even longer (77 days) when modeling the 
outburst decay \citep{deg11}.

\subsection{Data Reduction Procedure for Coherent Analysis}

A previous coherent timing analysis of \source is briefly described in
\citet{cav11}, who reported a pulse frequency derivative of
$\approx\,1.4\times\,10^{-12}\rm\,Hz\,s^{-1}$ with errors calculated
by means of Monte Carlo (MC) simulations that take into account the
red noise in the timing residuals \citep{har08}. The steps followed
are those used in standard coherent timing of radio and X-ray pulsars:
the photon arrival times are referred to the solar system barycenter
using the optical position from \citet{poo10}, fine clock corrections
are applied to the \textit{RXTE} data, and Earth occultations and intervals of
unstable pointing are filtered out.

To measure the TOAs of the persistent pulsations,
all bursts are removed from the light-curve, then $\approx10$ to
$\approx500$ s long intervals of data are folded in profiles of $N=32$
bins with the preliminary ephemeris reported in \citet{pap11}.  The
length of the folded data segments is chosen to have a nearly uniform
statistical error ($\sim 1$ms) for each TOA. The $\approx 10\rm\,s$
profiles are those folded from the first \textit{RXTE} data-set (ObsId
95437-01-01-00) when the pulse profiles have the highest fractional
amplitude and the highest signal-to-noise ratio. All other
observations are folded with $\approx 500$ s segments with the exact
length slightly varying for each observation to assure that all good
photons are retained in the data analysis. We assume that pulsations
are detected in a profile if the ratio between the amplitude of a
harmonic and its statistical error is larger than 3.5. This guarantees
that the expected number of false pulse detections is less than one in
the entire data collection (the total number of profiles is
$\approx650$ for each harmonic). A fundamental and two overtones are
detected with strong significance and in enough data segments to allow
a determination of a precise global timing solution.

The TOAs of each harmonic are fitted using the TEMPO2 pulse
timing program \citep{hob06}, assuming a circular Keplerian
orbit plus a pulse frequency and its first time derivative.  The entire
folding and fitting procedure is then repeated until the final timing
solution converges. The orbital parameters are reported in
Table~\ref{tab1} along with $1\,\sigma$ error bars obtained by
multiplying the statistical error as given by TEMPO2 by
$\sqrt{\chi^2}$. The errors on the pulse frequency and its derivative
are instead treated separately to include the strong effect of timing
noise which operates on the same timescales over which the two
parameters are measured. Details of the procedure have been
extensively discussed in the literature, and we refer to \citet{cav11}
for the results obtained on \source and to \citet{har08}
and \citet{pat10} for further details on the procedure.

\begin{table*}
\centering
\caption{Orbital Solution of \source}
\begin{tabular}{llll}
\hline
\hline
Orbital Parameter & Fundamental & First Overtone & Second Overtone\\
\hline
$P_{b}$  [hr]  & 21.27458(7) & 21.27470(7) & 21.2746(2)\\
$a_x\,sin(i)$ [lt-s] & 2.4965(4) & 2.4962(4) & 2.498(2)\\
$T_{asc}$ [MJD] & 55481.78025(4) & 55481.78015(5) & 55481.78018(2)\\
$e$ [$95\%$ c.l.]&  $<0.002$ & $<0.004$ & $<0.004$\\
\hline
\end{tabular}
\label{tab1}
\end{table*}

Here we will verify more carefully whether the pulse frequency
derivative (i.e., the measured $\dot{\nu}$) reported in \citet{cav11}
is indeed the spin frequency derivative (i.e., a true rotational
variation of the NS spin) and not the effect of timing noise.
Although \citet{cav11} used MC simulations to calculate the long-term
average $\dot{\nu}$ of \source, no interpretation to the pulse
frequency derivative is given and no discussion of the effect of
timing noise is provided.  Another long-term $\dot{\nu}$ has been
reported by \citet{pap11}. It is of the same order as, but
significantly different from,that given in \citet{cav11}.  The timing
analysis of \citet{pap11} did not consider in detail the strong effect
of timing noise and the authors reported discordant $\dot{\nu}$ values
in their analysis of the fundamental and first overtone which deviate
by more than 32$\sigma$ from each other and deviate by more than
$3\sigma$ from the results of \citet{cav11}. Identical considerations
also apply to the pulse frequency. These results show why extra care
has to be taken to verify that the pulse frequency derivative is the
spin frequency derivative of \source. Determining whether the NS is in
a spin-up phase is crucial for understanding the evolutionary history
of the binary as we will discuss in Section~\ref{evolution}.

We split the data in five segments of length between $\approx4$ and
$9$ days (see Table~\ref{tab:split}). The exact values are chosen empirically to have enough data
to measure a pulse frequency derivative of the order of
$10^{-12}\rm\,Hz\,s^{-1}$.  The data used and the start and end times of each
data segment are reported in Table~\ref{tab:split}. 
\begin{table*}
\centering
\caption{X-Ray Observations}
\begin{tabular}{lllll}
\hline
\hline
Segment Nr. & Obs Ids (95437) & Start & End & Flux\\
    & &  (MJD) & (MJD) & (Crab) \\
\hline
1 & 01-01-* 01-02-* \\
 & 01-03-* 01-04-* 01-05-*
\smallskip
\smallskip
  & 55482.006 & 55486.755 & 0.411 \\
2 & 01-06-* 01-07-* 01-08-*\\ 
&  01-09-* 01-10-00 01-10-01\\
\smallskip
\smallskip
& 01-10-02  & 55487.306 & 55493.347 & 0.499 \\
3 & 01-10-03 01-10-04 01-10-05\\ 
\smallskip
\smallskip
& 01-10-06 01-10-07 01-11-03 & 55494.246 & 55501.326 & 0.382 \\
4 & 01-11-04 01-11-05 01-11-06\\
\smallskip
\smallskip
  & 01-11-08 01-12-*    & 55502.407 & 55511.115 & 0.316 \\
5 & 01-13-* 01-14-*   & 55512.153 & 55519.165 & 0.259\\
\hline
\end{tabular}
\label{tab:split}
\end{table*}
We fit the pulse frequency and its first
derivative to each data segment. The fundamental frequency and the
first overtone are analyzed separately, whereas the second overtone is
not detected in enough observations to allow data splitting and 
will therefore not be consider any further. The harmonic phase
residuals relative to our best constant frequency timing model exhibit
variability on time scales as short as a few tens of minutes (i.e.,
close to the shortest timescale we are sensitive to) with an amplitude
well in excess of the Poisson noise expected from counting
statistics. This variability requires a special treatment of the
statistical errors of the pulsar spin parameters since standard
$\chi^{2}$ minimization techniques (i.e., those used by TEMPO2 or any
other standard fitting routine) rely on the assumption that only
Poissonian noise is present in the data, which is not a good
approximation in \source. We apply the MC method
developed by \citet{har08} to estimate the effect of timing noise on
the pulse parameters.  We run $10^{4}$ MC simulations on time series
which have the same power content of the original TOA residuals. The parameters
$\nu$ and $\dot{\nu}$ are measured for each simulated time series and
a distribution of parameters is constructed over the $10^4$
simulations. The $68\%$ confidence interval on the pulse parameters is
then determined as the standard deviation of each parameter
distribution.

A second test is performed by repeating the entire procedure described
above with the difference that we fit only a constant pulse frequency
with the derivative set to zero.  In this way we perform a ``partial
coherent'' analysis: coherent timing is used in each of the five data
segments to extract a constant pulse frequency (constant in that
segment), whereas the long term pulse frequency derivative is inferred
later by fitting the five pulse frequencies with standard $\chi^2$
minimization techniques. In this way we can check whether the five
constant pulse frequencies increase linearly in time as expected if a
constant spin frequency derivative is present in the data. If instead
the five frequencies fluctuate with a large scatter around a
linear trend, the effect of timing noise is predominant and a presence
of a constant spin frequency derivative is questionable.

To establish whether the pulse frequency
derivative is consistent with a spin frequency derivative we require that:
\begin{itemize}
\item the pulse frequency derivative inferred for the two harmonics is
  consistent with being the same and it also matches with the
  long-term pulse frequency derivatives measured with coherent timing
  (i.e., those published in Table 1 by \citealt{cav11});
\item the pulse frequencies have to be consistent with each other for the two harmonics in each of the five segments; 
\item there is no large deviation ($>3\sigma$) of any of the pulse
  frequencies from the linear trend. This demonstrates that timing
  noise is not dominating the fitted pulse frequency parameter in any
  of the five segments.
\end{itemize}
\begin{figure*}[t!]
  \begin{center}
    \rotatebox{0}{\includegraphics[width=1.0\textwidth]{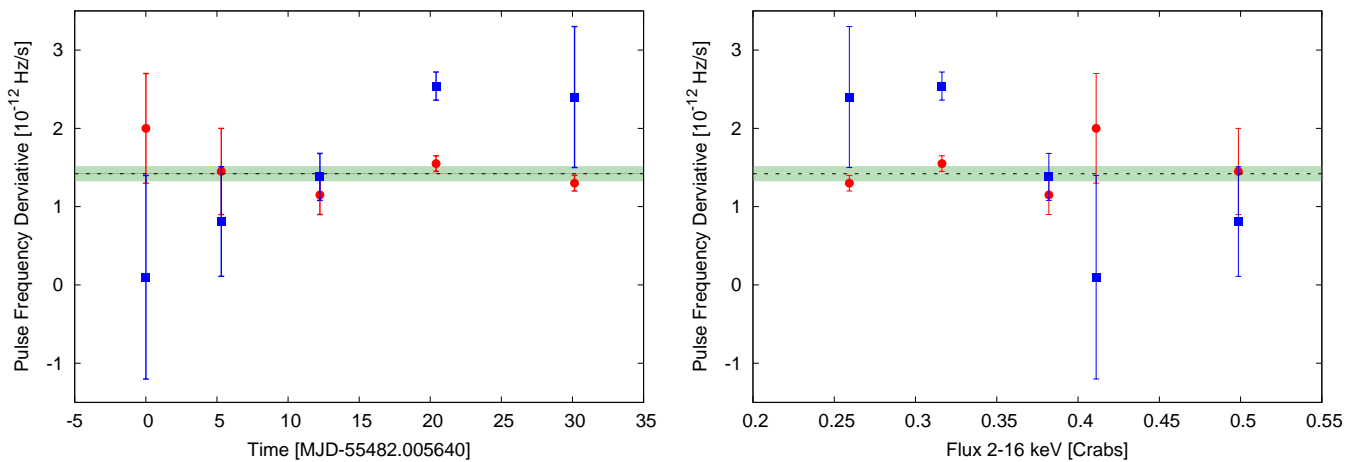}}
  \end{center}
  \caption{\textbf{Left Panel:} Time evolution of the pulse frequency
    derivative of \source for the fundamental (blue squares) and first
    overtone (red circles). The dotted line refers to the long-term
    average $\dot{\nu}$ of the fundamental reported in \citet{cav11},
    with the green band being the 68\% confidence interval calculated
    with MC simulations. Each data point is measured in a $\sim4-9$d
    data interval and the errors are calculated by means of MC
    simulations (see Section~\ref{xray}). All the data points are
    consistent with the green confidence interval bar with the
    exception of the 4th data point of the fundamental which can be
    considered an outlier (see Section~\ref{doesspinup} for a
    discussion). \textbf{Right Panel:} Pulse frequency derivative
    vs. X-ray flux. The meaning of symbols is the same as in the left
    panel. No trend is visible in the data suggesting that short-term
    $\dot{\nu}$ measurements have a very weak or no dependence on the
    X-ray flux.
    \label{torque}}
\end{figure*}

\section{Results of Coherent Analysis}\label{results}

We produce two different outputs: the pulse frequency derivative as it
evolves in time (Figure~\ref{torque}, left panel) and as a function of
the X-ray flux (Figure~\ref{torque}, right panel).  
The parameter shows no clear
evolution with time and the distribution of points around the long-term average
$\dot{\nu}$ is consistent with random fluctuations due to measurement
errors, with the exception of one outlier in the fourth data segment of the
fundamental.

Under the assumption that the X-ray flux tracks the mass accretion
rate $\dot{M}$ and that our five measured pulse frequency derivatives
represent short-term instantaneous spin frequency derivatives, we
expect to see $\dot{\nu}\propto\,f^{\gamma}_{X}$, where
$\gamma=\frac{6}{7}$ in the simplest accretion model (see for example
\citealt{gho79}, \citealt{bilchak97}).  From the right panel of
Figure~\ref{torque} we see no clear dependence of $\dot{\nu}$ from
$f_{X}$. All short term pulse frequency derivatives are consistent
within 1 $\sigma$ with the average long-term pulse frequency
derivative $\dot{\nu}=1.42(5)\times10^{-12}$, reported by
\citet{cav11}, except the second data point of the fundamental
frequency which deviates by more than 3$\sigma$ and is the same
outlier seen in the left panel of Figure~\ref{torque} (see
Section~\ref{doesspinup} for further discussion of the outlier).

\begin{figure}[t]
  \begin{center}
    \rotatebox{-90}{\includegraphics[width=0.65\columnwidth]{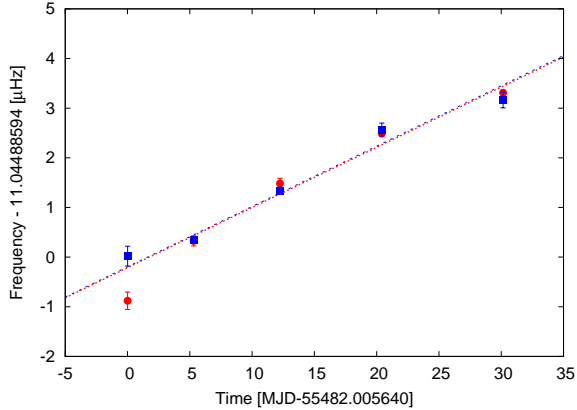}}
  \end{center}
  \caption{Pulse frequency evolution with time. The data points refer
    to the same data segments used in Figure~\ref{torque} and described
    in the text. The blue squares and red circles refer to fundamental
    and first overtone. Each pulse frequency is calculated with a
    constant pulse frequency model. Both data-sets (fundamental and
    first overtone) can be fit with a linear relation (blue dotted:
    fundamental, red dashed: first overtone) which is consistent with
    being the same within the statistical errors and gives a long-term
    $\dot{\nu}$ consistent with the value reported in \citet{cav11}.
    \label{torque2}}
\end{figure}

Finally, in Figure~\ref{torque2} the constant pulse frequencies are
plotted against time for fundamental and first overtone. We fit each
harmonic with a linear function $f(t)=k+\dot{\nu}\,t$ to extract the
long-term pulse frequency derivative.  The results give $\dot{\nu}$
and $k$ for the two harmonics which are consistent within the
$1\sigma$ statistical errors and are both consistent within 1$\sigma$
with the long-term pulse frequency derivative reported in
\citet{cav11}. 

\subsection{The Spin-up of \source}\label{doesspinup}

The results presented above strongly suggest that the long-term pulse
frequency derivative reported by \citet{cav11} is indeed the long term
spin frequency derivative of \source. There are however two points
that need further discussion before drawing a firm conclusion from our
data analysis: the meaning of the outlier in the short-term pulse
frequency derivatives and the lack of correlation between $\dot{\nu}$
and $f_{X}$.

\begin{figure*}[t]
  \begin{center}
    \rotatebox{0}{\includegraphics[width=0.90\textwidth]{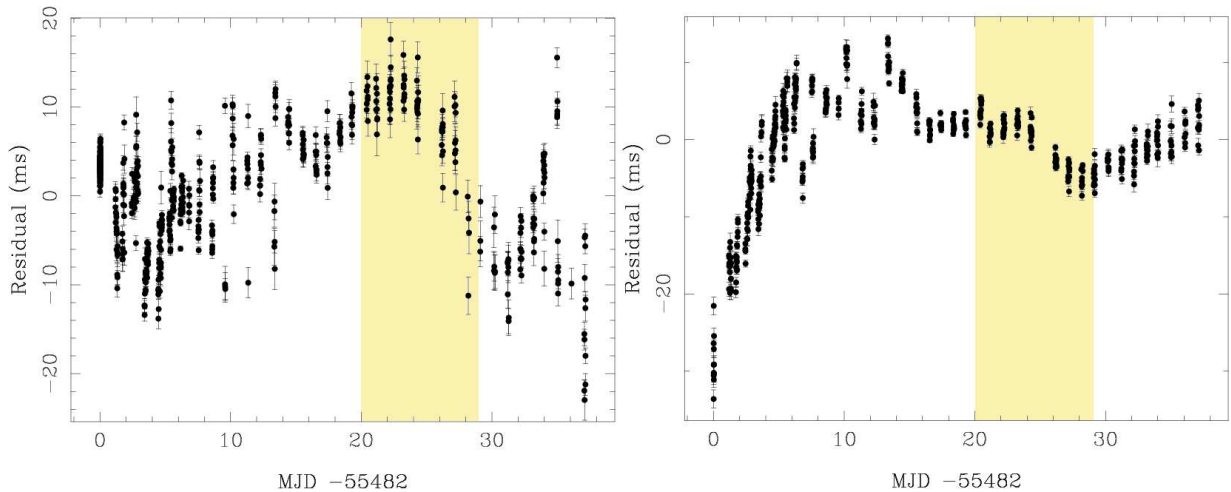}}
  \end{center}
  \caption{Timing residuals for the fundamental (left panel) and first
    overtone (right panel). The strength of timing noise for the two
    harmonics is comparable with an rms of the residuals for
    fundamental and first overtone of 5.1 and 4.9 ms respectively. The
    fundamental shows a stronger timing noise than the first overtone
    on short timescales ($\sim1$d) for the first 15 days and the last
    3 days, whereas the first overtone has a stronger noise component
    on longer timescales. The yellow shaded area refer to the data
    segment where the fundamental pulse frequency derivative shows an
    outlier (see text for a discussion) \label{res}}
\end{figure*}

Although the outlier indicates a significantly larger pulse frequency
derivative than the long-term average, the first overtone gives in the
same data segment a $\dot{\nu}$ incompatible with the outlier and
compatible with the long-term average $\dot{\nu}$. The interpretation
we give to this is that the outlier does not represent a true increase
in the spin frequency derivative of the NS. This can be still ascribed
to timing noise, even if we are carefully calculating the statistical
errors with MC simulations. MC simulations verify whether a measured
$\dot{\nu}$ is significant with respect to the red noise component in
the TOA residuals. This is achieved by testing whether random
fluctuations in the residuals are able to produce fake $\dot{\nu}$ as
strong as the measured one. Since we are looking at short time series
($\sim4-9$days), the red noise is not stationary, as evident from
Figure~\ref{res}, where the data segment to which the outlier belongs
is shaded. In that data segment, the highest red noise frequencies
observable are comparable with the length of the data segment and the
data show a simple parabolic trend which is absorbed by $\dot{\nu}$
during the fitting procedure. If this is the case, MC simulations
cannot discriminate between a true spin frequency derivative and the
effect of timing noise. Therefore the error bars
thus calculated do not differ much from those one
obtains when the errors are dominated by uncorrelated Gaussian noise
(i.e., white noise). The final values of the MC errors are
not, is such case, a good representation of the true statistical
uncertainties. This is a limitation of the MC method. However, a
similar behavior is not observed in the same data segment for the
first overtone and so we ascribe this behavior to timing noise.

It is of course possible that timing noise is the result of true
changes of the spin frequency and/or spin frequency derivative of the
NS. Fluctuating torques have been proposed in the past as a
mechanism to explain fluctuations in the NS rotation
\citep{lam78}. However, in this case we would expect to see identical
variations of the spin parameters in both harmonics which is not
observed for the outlier in \source. Furthermore it has been shown
that the timing noise observed in at least two accreting millisecond pulsars
\citep{wat08, pat10} cannot be ascribed to true variations of the neutron
star spin parameters since these require unphysical large torque
variations.

The fact that $\dot{\nu}$ is not proportional to $f_{X}$ can be a
consequence of different phenomena.  The X-ray flux might be not a
good tracer of $\dot{M}$ as has been already seen in several other
X-ray binaries \citep{van01}.  
Our data suggest that $\dot{\nu}\approx\rm\,const$ over
an X-ray flux variation of a factor $\approx\,2$. An approximately constant
 bolometric flux might therefore easily explain the observations.

Another possibility is that
$\dot{\nu}\propto\,f_{X}^{6/7}$ is valid only if the accretion disk is
a standard thin disk under the rather unrealistic assumption that
radiation pressure is not important in determining the inner disk
structure \citep{psa99}. \citet{and05} suggest that this assumption is not correct
already at accretion rates of a few percent Eddington. Since \source
is accreting at least at $\dot{M}>5\%$ Eddington with peaks exceeding 37\%, it
is not unreasonable to assume that radiation pressure plays a role in
determining the inner disk structure. The disk-magnetosphere
interaction region is also more complex than usually assumed (see
e.g.,~\citealt{rap04},~\citealt{dan10},~\citealt{kaj11},~\citealt{pat12}), and
$\dot{\nu}$ might have a weaker dependence on the bolometric flux than
usually assumed. 

We conclude therefore that the interpretation of the long term
$\dot{\nu}$ as being the spin frequency derivative of \source is
supported by the observations, with the outlier most likely explained
by unmodeled timing noise and the lack of correlation between the
short term $\dot{\nu}$ and $f_{X}$, which is not surprising, since the
expectation of such correlations is based on assumptions about the
accretion process which may not be valid.

\section{Evolutionary History of \source}\label{evolution}

The present RLOF spin-up epoch has been preceded by two earlier epochs
in which the binary was still detached. In the first dipole spin-down
epoch, the newly born NS has an initial rotation rate
$\Omega_0$ (spin $\nu_0=\Omega_0/2\pi$) and dipole moment $\mu_0$. The
magneto dipole torque dominates this epoch until the wind of the
companion penetrates the light cylinder and the NS is spun down by
wind torques. When this second epoch starts, the NS has rotational
rate and dipole moments $\Omega_1$ and $\mu_1$ respectively.  In this
section we present an evolutionary history of these three epochs to
determine the evolution of the NS spin in \source.  The dipole moment
decay induced by the spin-down of the NS in the
superfluid-superconducting core of the star is assumed to be an
essential factor during the spin-down epochs.  We summarize the
duration and the symbols used for each evolutionary phase in
Table~\ref{tab2}.

\begin{table}
\centering
\caption{Evolutionary History of \source}
\begin{tabular}{lccccc}
\hline
\hline
Epoch & Duration (yr)& $\mu_i$ & $\mu_f$ & $\Omega_i$ & $\Omega_f$ \\
\hline
Dipole spin-down & $10^7$ & $\mu_0$ & $\mu_1=\mu_0$ & $\Omega_0$ & $\Omega_1$ \\
Wind spin-down & $10^7-10^8$ & $\mu_1=\mu_0$ & $\mu_2$ & $\Omega_1$ & $\Omega_2$ \\
RLOF spin-up & $10^7-10^8$ & $\mu_2$ & $\mu_2$ & $\Omega_2$ & $\Omega_3$ \\
\hline
\end{tabular}\label{tab2}
\tablecomments{The subscripts $i$ and $f$ refer to the initial and final value at a specific epoch. 
  The magnetic moment $\mu_2$ is assumed to remain constant during the RLOF phase.}
\end{table}

\subsection{Duration of the Roche Lobe Overflow Epoch}

In the present RLOF spin-up epoch, the 2-16 keV X-ray luminosity of
\source spans values between $9\times10^{36}\rm\,erg\,s^{-1}$ and
$6.5\times10^{37}\rm\,erg\,s^{-1}$ for an assumed distance of 5.5 kpc
\citep{ort07}. 
This luminosity has to be considered a lower
limit given that the bolometric luminosity is expected to be higher
than the X-ray luminosity. \citet{lin12} estimated a bolometric
luminosity at the peak of the outburst of about $50\%$ the Eddington
limit for $d=6.3$ kpc ($L_{Edd}=2.5\times10^{38}\rm\,erg\,s^{-1}$).
The maximum mass accretion rate is therefore

\begin{equation}\label{mdot}
\Mdot = \frac{LR_{NS}}{GM_{NS}} \simeq 5 \times 10^{17} \left(\frac{d}{5.5{\rm\,kpc}}\right)^2
\frac{R_{6}}{M_{1.4}}\rm\,g\, s^{-1}, \nonumber
\end{equation}

where $G$ is the gravitational constant, $M_{NS}$ and $R_{NS}$ are the
mass and radius of the NS, and $M_{1.4}$ and $R_{6}$ are the mass of the
NS in units of $1.4\msun$ and the radius in units of $10^6\rm\,cm$
respectively. The average mass accretion rate during the outburst has
a value approximately half the peak value, and we assume it is
$\langle\,\Mdot\,\rangle\,= 2\times10^{17}\rm\,g\,s^{-1}$, in agreement
with the value reported by \citet{deg11}.

The inner radius $r_A$ of the accretion disk is determined by the
balance between the magnetic stresses in the NS
magnetosphere and the material stresses of the accretion disk
\citep{gho79}:

\begin{equation}\label{ra}
r_{A}= \left(\frac{{\mu}^2}{\Mdot}\right)^{2/7}\left(GM_{NS}\right)^{-1/7}.
\end{equation}

where $\mu$ is the dipole magnetic moment of the star. The Keplerian
rotation rate at the inner radius of the disk

\begin{equation}
\Omega_K(r_A) = \left(\frac{GM_{NS}}{{r_A}^3}\right)^{1/2} =
\left(\frac{{\mu}^2}{\Mdot}\right)^{-3/7}\left(GM_{NS}\right)^{5/7}
\end{equation}

determines the torques on the NS applied by the accretion flow and the
induced distortions of the magnetic field.  These torques operate as
long as $\Omega_{\*}$ is not equal to $\Omega_{K}(r_A)$, so that the
star spins up or down toward an equilibrium rotation rate
$\Omega_{eq} = \Omega_K(r_A)$. If any of the QPO frequencies (see
\citealt{alt10}) is interpreted as the eventual spin equilibrium
frequency of the NS, the present rotation rate will give an estimate
of the dipole moment of the NS. This identification however, is not
straightforward. QPO frequencies reflect modes of oscillation in the
accretion flow that are excited in the interacting disk-NS system to
some amplitude to effect the luminosity signal. A particular QPO
frequency band might be defined by some oscillation mode of the disk
or by a beat or resonance between the NS and a disk mode. Disk modes
are fundamentally effected by rotation and imprinted by the frequency
scale of the disk rotation, Keplerian in most parts of the disk, but
necessarily and significantly deviating from the Keplerian rate at the
inner boundary where the effect of the NS magnetosphere is
important. This results in frequencies of epicyclic (radial) disk
oscillations to be somewhat higher than the non-Keplerian rotation
frequency at the inner disk boundary or transition region
\citep{alp08}.  The presence of a sequence of QPO frequencies may be
due to the presence of several multipole components of the neutron
star magnetic field, so that the boundary between the disk and the NS
magnetosphere is convoluted with a sequence of decreasing ``stopping
radii'', smaller than the conventional Alfven radius given in
Equation (\ref{ra}) and increasing {\it{Kepler}} frequencies corresponding to
higher multipoles of the NS magnetic field (\citealt{alp11}). We assume
that one of the observed range of QPO frequencies $\nu_i$ is close to
the Keplerian rotation rate $\Omega_K(r_A)$ near the Alfv\'{e}n radius,
\begin{equation}
\nu_i \cong \frac{\Omega_K(r_A)}{2\pi} \nonumber
\end{equation}
yielding the estimates
\begin{eqnarray}
\mu & \cong & (2\pi \nu_i)^{-7/6} \dot{M}^{1/2} (GM_{NS})^{5/6}\\\label{mustart}
\mu & = & 1.3 \times 10^{26}\rm\,G\; {cm}^3\left(\frac{\nu_i}{kHz}\right)^{-7/6}\avmdot^{1/2} M_{1.4}^{5/6} \nonumber
\end{eqnarray}
for the dipole moment. In these estimates we use $\avmdot$ to mean the
mass accretion rate averaged over the outburst length 
and it is normalized to $2\times10^{17}\rm\,g\,\; s^{-1}$; 

\begin{table}
\centering
\caption{QPO Frequency Model for \source}
\begin{tabular}{ccccc}
\hline
\hline
QPO Frequency $\nu_{i}$& $\mu$ & $B$ Field & $\dot{\nu}$ & $t_{su,QPO}$\\
$(\rm\,Hz)$ & $(10^{27}\rm\,G\,cm^3)$ &  ($10^{9}$G) & $(10^{-12})$ & $(10^7\rm\,yr)$\\
\hline
48 & 5 & 10 & 1.6 & 1.8\\
173 & 1 & 2 & 1.1 & 2.8\\
814 & 0.2 & 0.4 & 0.6 & 4.6\\
\hline
\end{tabular}\label{tab3}
\tablecomments{The magnetic field $B$ refers to the value 
at the NS magnetic poles. This is twice the value at the equator.}
\end{table}

The values of the magnetic dipole moment are reported in Table~\ref{tab3}. 
If we take the NS to be already in equilibrium at the 
present spin frequency of 11 Hz, then $\mu \cong 2 \times 10^{28}\rm\,G\,{cm}^3 $
and a $B$ field at the poles of $4\times10^{10}$G where we have used:
\begin{equation}
\mu_{30} = 0.6 f^{7/6}\left(\frac{d}{5.5{\rm\,kpc}}\right) {R_{6}}^{1/2} M_{1.4}^{1/3}\nu^{-7/6}
\end{equation} 
Here $f$ takes account of the possibility that the observed frequency
is not the {\it{Kepler}} frequency but some related frequency at the disk
boundary: e.g., $f \sim 1.5-2$ for the epicyclic frequency (e.g.,
\citealt{alp08}). This uncertainty makes the field
estimates somewhat higher.  \\ The torque applied by the accretion
disk on the NS is usually assumed to be:
\begin{equation}
N = \Mdot (G M_{NS} r_A)^{1/2} F(\omega)
\end{equation}
with the dimensionless torque $F(\omega)$ describing the
dependence on the fastness parameter $\omega \equiv
\Omega_{\*}/\Omega_K(r_A)$, is likely of the form
\begin{equation}
F(\omega) = 1 - \omega^p \nonumber
\end{equation}
with the exponent $p \sim 2$ \citep{ert09}. If \source is as
yet at the beginning of its spin-up epoch, then $\omega << 1$,
$F(\omega) \cong 1$, giving constant $\nudot$ and expected spin
frequency derivatives
\begin{equation}
\dot{\nu} \cong 6 \times 10^{-13}{\rm\,Hz\; s^{-1}}  \frac{\avmdot M_{1.4}^{2/3}}{ I_{45}\left(\frac{\nu_i}{\rm\,kHz}\right)^{1/3}}
\end{equation}
where $I_{45}$ is the NS moment of inertia in units of $10^{45}\rm\,g\,cm^2$. These estimates of
$\dot{\nu}$ are also given in Table~\ref{tab3}. With any of the QPO $\nu_i$
this gives $\dot{\nu} \sim 10^{-12}\rm\,Hz\; s^{-1}$.  If the system is
already in equilibrium $\dot{\nu}$ would be fluctuating in sign with
rms values much smaller than $10^{-12}\rm\,Hz\; s^{-1}$, which is however
not observed (see Section 3 and Figures~\ref{torque} and ~\ref{torque2}).

The spin-up age up to the present is then
\begin{equation}\label{spinupQPO}
t_{su,QPO} \cong  5\times10^5 \;yr \frac{ I_{45} {\left(\frac{\nu_i}{kHz}\right)}^{1/3}}{\avmdot M_{1.4}^{2/3}\Delta}.
\end{equation}
where $\Delta$ is the (unknown) duty cycle of the binary.  If we
assume $\Delta\sim0.01$, the maximum estimated time in
the spin-up epoch up to now is $\approx 5\times10^7$ yr corresponding to
$\nu_i = 814 \rm\,Hz$ (see Table~\ref{tab3}).  If any of the QPO
frequencies is interpreted as the eventual equilibrium frequency, then
the time spent in the current spin-up epoch is a small fraction of the
age of the system.  Even if the system were already in equilibrium at
$\nu = 11\rm\,Hz$ then it must have reached equilibrium in the short
time
\begin{equation}
t_{su} \cong \frac{\nu}{\dot{\nu}\,\Delta}\simeq 2.5\times10^7 yr. 
\end{equation}

\subsection{Duration of the Dipole Spin-down Epoch} 

 New born NSs have a distribution of rotation rates $\Omega_0 \sim
 1-300$ rad s$^{-1}$ and dipole magnetic moments $\mu_0 \sim 10^{29} -
 10^{30}$ G cm$^3$ \citep{fau06}. The NS is active as a pulsar until
 the magnetospheric voltage falls below a critical threshold (the
 ``pulsar death line''). For conventional radio pulsars the epoch of
 pulsar activity lasts for 10$^6$-10$^7$ yr. No evidence of magnetic
 field decay is seen in the pulsar population during this active
 epoch. The lower limit for an exponential decay time is 10$^7$ yr.

If the magnetic dipole moment decayed in proportion to the rotation
rate, due to flux-line vortex line coupling \citep{sri90},
\begin{equation}\label{mu}
\mu(t) = \mu_0 \frac{\Omega(t)}{\Omega_0}.
\end{equation}  
the spin-down would follow a power law with braking index $\cong
5$. More detailed considerations of flux-line motion indicates that
the simple scaling in Equation (12) is a likely to be a good description
for the evolution~\citep{jah00}.

The vortex-flux coupling leads to the decay of magnetic flux in the
NS superfluid neutron-superconducting proton core. For the
total dipole moment to decay, the flux expelled from the core must
diffuse through the highly conducting crust. The decay time through
the crust is estimated to be 10$^7$ yr. This is supported by the
limits on exponential decay of the dipole moment from population
analysis of the radio pulsars. Ongoing dipole spin-down on time scales
much longer than 10$^7$ yr, that is, beyond the typical duration of
the epoch of radio pulsar emission, will be effected by the power law
decay of the dipole moment due to flux-vortex coupling.

Whether pulsar activity has ceased or not, vacuum dipole spin-down will
continue until the wind captured from the companion penetrates the
light cylinder. The captured wind loses angular momentum in a shock
and will accrete towards the NS down to a stopping radius
$\sim r_A$. The wind will be able to affect the torque on the neutron
star when
\begin{equation}\label{windt}
r_A (\mu_0, \dot{M}_{wNS}) \simless r_{LC} \equiv \frac{c}{\Omega_1}.
\end{equation}
with $ \dot{M}_{wNS}$ denoting the wind \textit{captured} by the NS. 
After this point, spin-down will proceed under the effect of the inflowing wind.  From Equation (\ref{windt}) the rotation rate
 $\Omega_1$ at the beginning of the wind spin-down epoch is
\begin{equation}\label{omega1}
\Omega_1 \cong 24 {\rm\,rad\; s^{-1}}  \mu_{1,29}^{- 4/7} \dot{M}_{wNS,13}^{2/7}  M_{1.4}^{1/7} 
\end{equation}
where $\dot{M}_{wNS,13}$ is the mass inflow rate arriving at $r_A$ in
units of $10^{13}$\rm\,g\, s$^{-1}$.  If we first assume that
flux-vortex coupling has no effect, the dipole spin-down proceeds with
a constant dipole moment and braking index 3:
\begin{equation}\label{nudot}
\dot{\Omega} = -K \Omega^{3} = -\frac{4 \mu^2 \pi}{3 c^3 I} \Omega^3
\end{equation}
where $c$ is the speed of light and $I$ is the NS moment of
inertia.  The duration of the dipole spin-down epoch, until the
rotation rate $\Omega_1$ is reached is given by
\begin{eqnarray}
t_{n=3} & \equiv & \frac {P_1}{2\dot{P}}=-\frac{\Omega_1}{2\dot{\Omega}} \nonumber \\ 
& \cong & 10^{10} yr\; \mu_{1,29}^{-2}\Omega_1^{-2} \nonumber \\
& \cong & 2.5\times 10^6 yr\; \mu_{0,30}^{- 6/7} \dot{M}_{wNS,13}^{-4/7}  M_{1.4}^{-2/7} \label{n3}
\end{eqnarray}
where we have used Equations (\ref{omega1}) and (\ref{nudot}) and
$I=10^{45}\rm\,g\,cm^2$.  In the last line of Equation (\ref{n3}) we have
assumed that the magnetic field does not decay during the dipole
spin-down epoch, so that the dipole moment $\mu_1$ at the end of this
epoch is still the dipole moment $\mu_0$ of the NS at
birth. Since the lifetime thus obtained is not much smaller than the crust
diffusion timescale of 10$^7$ yr, it might be necessary to take the 
vortex-flux coupling and the induced decay of the dipole magnetic
moment into account. The correct estimate of dipole spin-down epoch, incorporating
flux decay according to Equation (\ref{mu}) proceeds with braking index
5. We obtain
\begin{eqnarray}
t_{n=5} & \equiv & \frac {P_1}{4\dot{P}} = -\frac{\Omega_1}{4\dot{\Omega}}\nonumber \\ 
& \cong & 5 \times 10^{9} yr\; \mu_{1,29}^{-2}\Omega_1^{-4} \nonumber \\
& \cong & 1.5 \times 10^4 yr\; \mu_{1,29}^{2/7} \dot{M}_{wNS,13}^{-8/7}  M_{1.4}^{-4/7}. \nonumber \\
\end{eqnarray}
Relating the magnetic moment $\mu_1$ and rotation rate $\Omega_1$ at the end of dipole spin-down to the initial values $\mu_0$ and $\Omega_0$ with Equation (\ref{mu})
\begin{equation}\label{mu1}
\mu_{1,29} = 10 \mu_{0, 30} \frac{\Omega_1}{\Omega_0}, 
\end{equation}
and using Equation (\ref{omega1}) we obtain  
\begin{equation}\label{n5}
t_{n=5} \cong 2 \times 10^4 y\;\mu_{0,30}^{2/11}\left[\frac{\Omega_0}{10{\rm\,rad\,s^{-1}}}\right]^{-2/11} \dot{M}_{wNS,13}^{-12/11}  M_{1.4}^{-6/11}.
\end{equation}
This value is shorter than the typical field diffusion time scale
through the NS crust ($\simgreat 10^7$yr) and therefore
Equation (\ref{n5}) cannot be used self-consistently for $\mu_0\sim
10^{29}-10^{30}\rm\,G\,cm^3$.
If one ignores the field decay and assumes a typical
$\mu_0=10^{29-30}\rm\,G\,cm^3$ in Equation (\ref{n3}) then the dipole spin down age is
$\sim10^7$ yr.

\subsection{Duration of the Wind Accretion Spin-down Epoch} 

Wind accretion
(\citealt{edg04},~\citealt{the93,the96},~\citealt{pfa02}) proceeds by
Bondi-Hoyle like capture of material from the companion's wind, which
is then channeled toward the NS by outward transport of angular
momentum in traversing a shock front. The wind mass loss rate from a
$\sim1\msun$ main sequence companion is $\Mdot_{wind} \sim
10^{12}\rm\,g\,s^{-1}$, which is low when compared to the present day
mass accretion rate, driven by RLOF. Even considering a sub-giant
phase of a $\sim1\msun$ donor star, the wind loss rate will increase
only slightly, reaching $\Mdot_{wind} \sim 10^{13}\rm\,g\,s^{-1}$ for
sub-giant stars (see for example \citealt{rei75,rei78}).  Only a
fraction of this wind will be captured to flow towards the NS, with
$\Mdot/\Mdot_{wind}$ as low as $\sim 10^{-2}$ \citep{nag04}. However,
since the orbit of \source shows stringent upper limits on the
eccentricity ($e<10^{-3}$, \citealt{pap11}) tidal circularization has
probably already taken place and it is likely that the donor star is
rotating synchronously with the orbit, since the circularization
timescale is typically larger than the synchronization timescale (see
for example \citealt{hur02}). The large rotation of the donor
($v_{rot}\simeq70\rm\,km\,s^{-1}$) can boost the wind loss rate by a
large factor of the order of $10^{2-3}$ when compared with typical
isolated stars of similar mass at the same evolutionary stage. The
captured wind might be therefore comparably higher, with values of
$\dot{M}_{wNS}\sim10^{13-14}\rm\,g\,s^{-1}$.

The NS is spinning rapidly within this wind of low specific angular
momentum. For our purposes of estimating the duration of the spin-down
epoch it will be enough to take a constant torque, giving the
spin-down rate:
\begin{eqnarray}
\dot{\Omega} & = & - \frac{\dot{M}_{wNS}}{I} \left(G M_{NS} r_A\right)^{1/2} \nonumber\\
& = & - \left(\dot{M}_{wNS}\right)^{6/7}\frac{\mu^{2/7}}{I}\left(G M_{NS}\right)^{3/7}\\
& = & -4.9\times10^{-15}{\rm\,rad\,s^{-1}}\dot{M}_{wNS,13}^{6/7}\mu_{1,29}^{2/7}M_{1.4}^{3/7}I_{45}^{-1}
\end{eqnarray}

Using Equations (\ref{mu}) and (\ref{mu1}) one can
find the spin-down in wind and field decay solutions
\begin{eqnarray}
{\mu(t)}^{5/7} & = & {\mu_1}^{5/7} - 5/7 C' t  \\
{\Omega(t)}^{5/7} & = & {\Omega_1}^{5/7} - 5/7 C t \\
C & \equiv & \left(\frac{\mu_1}{\Omega_1}\right)^{2/7}\frac{{\dot{M}_{wNS}}^{6/7}(GM_{NS})^{3/7}}{I} \nonumber \\
C' & \equiv & \frac{\mu_1}{\Omega_1}\frac{{\dot{M}_{wNS}}^{6/7}(GM_{NS})^{3/7}}{I}. \nonumber
\end{eqnarray}
\\ Since the magnetic moment does not change after the wind spin-down
epoch, $\mu_2$ at the end of the wind spin-down epoch is just the
magnetic moment in the present RLOF spin-up epoch as estimated above. 
By assuming that $\Omega(t_{sd}) = \Omega_2 <<\Omega_1$ 
and $\mu_2<<\mu_1$, we obtain
\begin{eqnarray}\label{sd}
t_{sd} & = & \frac{7}{5} \frac{\Omega_1}{{\mu_1}^{2/7}}
\frac{I}{{\dot{M}_{wNS}}^{6/7}(GM_{NS})^{3/7}} \nonumber \\
& \cong &  3\times 10^{7} \; yr \;I_{45}\;\mu_{0,30}^{-6/7}\dot{M}_{wNS,13}^{- 4/7}  M_{1.4}^{-2/7}.
\nonumber \\ 
\end{eqnarray}

where we have assumed that $\mu_1=\mu_0$ as found
at the end of the previous section.

From the calculations presented in Equation (\ref{sd}) we see that for a
captured wind of
$\dot{M}_{wind,NS}\sim\,10^{-2}\cdot\,\dot{M}_{wind}\sim10^{13}-10^{14}\rm\,g\,s^{-1}$,
the spin-down epoch lasts for a time $10^{7-8}$yr, much shorter than
the age of the cluster ($\sim10^{10}$yr) for an assumed
$\mu_1=\mu_0=10^{29-30}\rm\,G\,cm^3$. If more wind is accreted, as
described for example in the focused accretion model of \citet{pod07}
, the wind spin-down epoch would be further
shortened if the wind still carries low angular momentum. If the wind
has instead large angular momentum because it is focused through the
Lagrangian point $L_1$, the RLOF epoch would be further shortened.

\subsection{Incompatibility with a Spin Equilibrium Scenario}

In the previous section we have seen that \source is not close to
equilibrium since it exhibits a strong spin-up during the
outburst. Several other considerations on the evolutionary history of
\source also suggest that the NS is currently spinning up.  According
to the spin evolution history, the present 11 Hz spin frequency does
not represent a special frequency. It remains to be explained
whether the conclusion that the binary has currently completed only the first few million years since the onset of the RLOF, a phase which
lasts for $\sim10^9$ yr, requires any special pleading.

If we assume that the observed spin-up and outburst duration represent
the typical values of an outburst of \source, then the pulsar spins up
by $\dot{\nu}\;\Delta\,t_{o}\simeq 7\times10^{-6}\rm\,Hz$ at each
outburst, where $\Delta\,t_o$ is the outburst length (55 days).  If the
NS is dominated by magneto dipole spin-down during quiescence, as
proposed for some accreting millisecond pulsars (\citealt{har08,
  har09, har11}, \citealt{pat10}, \citealt{pap11}, \citealt{rig11})
then a balance of the spin-up requires a magnetic field
$B\simeq10^{11}\rm\,G$ and a recurrence time of $\sim15$ yr to provide
a spin-down with the necessary magnitude calculated via the expression
\citep{spi06}:
\begin{equation}
\dot{\nu}_{sd}=-\frac{\mu^2\left(\frac{2\pi\nu}{c}\right)^3\left(1+{\rm\,sin}^2\alpha\right)}{2\pi\,I}
\end{equation}
where $\alpha$ is the offset angle between the rotational and magnetic
field axes. The spin down would then be of the order of
$-10^{-14}\rm\,Hz\,s^{-1}$.  

However, as discussed in Section~\ref{intro}, different estimates of the
$B$ field of \source suggest $B\sim10^9-10^{10}$G. Such a $B$ value
requires a recurrence time of $\sim10^2-10^3$ yr which 
is not compatible with a binary evolution scenario with the donor
 in an RLOF phase. The mass transfer rate (equal to the 
outburst mass accretion rate averaged over the source duty cycle) would then
average to $\Mdot\sim10^{13}-10^{14}\rm\,g\,s^{-1}$, a value
too small for an evolved donor or even a main sequence star with
$M>0.8\msun$ (typical values being $\sim10^{15}-10^{16}\rm\,g\,s^{-1}$).

\section{Constraints on the Donor Mass and Binary Parameters}\label{constraints}

The mass function of \source has been reported in \citet{pap11} and
corresponds to a minimum donor mass of 0.4$M_{\odot}$ for a neutron
star of $1.4\msun$.  The location of \source in the globular cluster
Terzan 5 allows to place further constraints on the donor mass.
Terzan 5 is composed by two populations of stars: one with sub-solar
metallicity ($Z=0.01$, $Y=0.26$) and one with supra-solar metallicity
($Z=0.03$, $Y=0.29$; see~\citealt{fer09}). One interpretation of these
discrepant metallicities and Helium abundances is that the cluster is
composed by two different populations of stars with ages of $12\pm1$
Gyr (metal poor) and $6\pm2$ Gyr (metal rich).  If we assume that the
donor star belongs to the first population, strong constraints can be
placed on the donor star.  Since the orbital period of \source is
$\approx 21$ hr, and assuming a minimum NS mass of 1.2$\msun$, the
minimum orbital separation of the system is $A\approx0.02$ AU. By
using the Roche lobe approximation of \citet{egg83}:
\begin{equation}
  R_{L}=\frac{0.49\,A\,q^{2/3}}{0.6q^{2/3}+\rm\,ln\left(1+q^{1/3}\right)}
\end{equation}
with $q=M_{d}/M_{NS}$ being the ratio of the donor and accretor mass, the
minimum possible Roche lobe radius that corresponds to the minimum orbital
separation and the minimum mass ratio ($M_d=0.4\msun$, $M_{NS}=1.2\msun$)
is 1.1$R_{\odot}$.  After 12 Gyr all stars with a mass $M\simgreat
1\msun$ have evolved to the red giant phase and might have already turned into
white dwarfs whereas all stars with  $M\simless0.9\msun$ (i.e., the turn off mass of the metal-poor population of Terzan 5, \citealt{fer09}) are still on the main sequence.
Considering the other extreme case of $M_d=1.0\msun$ and $M_{NS}=2.0\msun$, 
the maximum Roche lobe radius will be 1.8$R_{\odot}$.

To fill the Roche lobe therefore any possible donor star of \source has to be
an evolved star that has left the main sequence and has increased its
radius to fit its Roche lobe. The donor star
mass falls in the range $0.9\simless\,M\simless\,1\,\msun$
and is likely to be a sub-giant. This result strongly
constrains the orbital separation of the system to be in the range
$0.023$-$0.026$ AU for a total mass of the binary between 2.1
($1.2\msun$ NS and 0.9$\msun$ donor) and 3$\msun$ ($2\msun$ NS and
$1\msun$ donor).  The inclination $i$ of the binary is then
constrained by using the projected semi-major axis of \source (see
Table~\ref{tab1}): 
\begin{equation}
i={\rm\,sin^{-1}}\left(\frac{5\times10^{-3}\rm\,AU}{A}\frac{M_{NS}+M_d}{M_{d}}\right)\simeq\,30^{\circ}.
\end{equation}
Identical considerations
apply if the donor star belongs to
the metal rich population, with the
difference that the turn off mass will be higher by a few tenths of
solar mass and the inclination smaller by a few degrees. 

All these considerations assume that irradiation of the donor star is
unimportant in determining its radius. This is a good assumption
given that non-negligible irradiation sets in with the onset of RLOF
and we have seen that this phase has lasted for $\simless10^7-10^8$
yr. Therefore any possible donor star has filled the Roche lobe 
following a standard evolutionary path that does not involve irradiation
of the outer envelope. The 
effect of irradiation might instead be important during the RLOF phase
when the outermost layers of the envelope expand after absorbing
 the energy deposited by the X-ray radiation. The star being
 constrained to fit the Roche lobe will increase its mass loss through
the inner Lagrangian point and boost the mass transfer rate.
The average X-ray power absorbed by the companion during an outburst is 
\begin{equation}
L_{abs}=\langle\,L_{x}\rangle\left(\frac{\,R}{2\,A}\right)^2
\end{equation}
with
$\langle\,L_{x}\rangle\sim4\times10^{37}\rm\,erg\,s^{-1}$
being the average X-ray luminosity during an outburst. Given a duty
cycle of $\Delta\simless0.005$, the long term average power absorbed
by the donor will be $\simless\,10^{33}-10^{34}\rm\,erg\,s^{-1}$. This value
corresponds to $\simless10\%-100\%$ of the outgoing power produced by a
low mass stars in the sub-giant branch and cannot be considered
negligible. The quantitative effect of irradiation can be calculated
with stellar evolutionary codes and is beyond the scope of this work.

\section{Discussion}\label{discussion}

We have investigated the evolutionary history of \source discussing
three distinct evolutionary epochs: a dipole dominated spin-down
epoch, a wind epoch and the currently observed RLOF phase. By taking
into account the effect of timing noise in the X-ray pulsations we are
able to confirm the long-term average spin-up of
$1.4\times10^{-12}\rm\,Hz\,s^{-1}$ observed during the 2010 outburst
(\citealt{pap11,cav11}). If the X-ray flux represents most of the
bolometric flux, the short term spin frequency derivatives $\dot{\nu}$
do not scale with the X-ray flux in the expected way. The long-term
$\dot{\nu}$, however, leads to exclude a scenario in which \source has
already reached spin equilibrium. A long-term spin equilibrium with
magnetic dipole spin down in quiescence balancing the accretion
torques in outburst is also excluded based on binary evolution
considerations.  The spin-up timescale suggests that \source has been
spun-up in the RLOF phase for a few $10^{7}$ yr. An interpretation of
the observed QPOs as the equilibrium frequencies of \source indicates
a similar timescale for the spin-up epoch up to the present.

The spin-down epochs suggest that the NS in \source had reached a
spin frequency of the order of $\sim1$Hz at the onset of the RLOF
phase, if we assume that the newborn NSs had typical initial
values for $\Omega_0$ and $\mu_0$.  This reinforces the idea that the
time elapsed in RLOF has been very short, since the time required to
spin-up a 1 Hz NS to 11 Hz with a spin frequency derivative of
$10^{-12}\rm\,Hz\,s^{-1}$ and a duty cycle of 0.01 is a few 
$10^7$yr.

Based on these findings we conclude that \source is in an
exceptionally early RLOF phase. The total spin-up timescale to
transform \source from a slow pulsar ($\sim\,1\rm\,Hz$) into a
millisecond one ($\nu>100\rm\,Hz$) is $\simgreat10^8-10^{9}$yr.  Since
the current RLOF phase will last for $\sim10^9$yr, we would expect
to observe today $\sim1-10$ accreting millisecond pulsars that have
followed a similar evolutionary history as \source. The accreting
millisecond pulsar SAX J1748.9-2021 ($\nu\simeq442\rm\,Hz$) in the
globular cluster NGC 6440 has a companion with mass and orbital period
compatible with being in a slightly evolved post main sequence phase
\citep{alt08}. It is likely that the companion of SAX J1748.9-2021
has followed an evolutionary history similar to \source with the
difference that it spent a longer time in the RLOF phase thus turning
its NS into a millisecond pulsar. The scenario outlined above is
compatible with the observed population of accreting pulsars, although
it is difficult to assess the problem in a robust way given the small
number statistics. It is also possible that dynamical interactions
have played a role by decreasing the number of similar systems in
globular clusters via ionization interactions. This however requires
further investigations along with detailed binary evolution
calculations.

The prior spin-down history gives results which are not compatible
with a binary having an age of several $10^9$ yr as expected if \source
is primordial or if it has not suffered exchange interactions in the
last few $10^9$ yr. Different formation scenarios are available to
explain the apparent discrepancy between the age of the cluster and
the binary.  A first possibility is that a recent dynamical encounter
has played a role in forming the binary or in accelerating the onset
of the RLOF phase. If an exchange interaction has taken place in the
last few $10^7-10^8$ yr then this would explain why the apparent age
of \source is so short compared with the age of the cluster. The
location of \source in the high mass and high central density cluster
Terzan 5 suggests that the interaction rate might be particularly high
in this environment. Dynamical simulations to calculate the typical
interaction rates in the core of Terzan 5 are required to assess this
question.

A second possibility for the origin of \source is formation of the
accreting pulsar via accretion induced collapse (AIC) of a massive
white dwarf (\citealt{miy80}, \citealt{nom87}). In this scenario a
ONeMg white dwarf accretes matter until its mass exceeds the
Chandrasekhar limit and it collapses to form an NS via
electron captures on Mg and Ne nuclei. In this case the binary must
have gone through a preliminary contact phase during which the white
dwarf was accreting from a donor star in RLOF. During
the collapse approximately 0.2$\msun$ are ejected from the binary (see
for example \citealt{van11} for a discussion) which causes a sudden
expansion of the orbit turning the system into a detached binary. At
this point the binary follows the three evolutionary epochs described
in Section~\ref{evolution} with the NS age set by the onset
of AIC.  The formation of NSs from AIC has been discussed by
\citet{lyn96} to explain the young radio pulsar population in globular
clusters. Lyne et al. note that the formation rate of young pulsars
in globular clusters is larger than that of millisecond pulsars in
globular clusters. The young pulsars are found in metal rich globular
clusters with high mass and high central densities, like Terzan 5
(\citealt{lyn96,boy11}). Further evidence that globular clusters might
contain young NSs was also presented by \citet{fre11}, who
reported the discovery of the youngest millisecond pulsar ever found
in the globular cluster NGC 6624, with a characteristic age $t<24$ Myr
(see also \citealt{lyn87} and \citealt{cog96} on the ``young''
millisecond pulsar PSR 1821-24A in the globular cluster M28).  We can
therefore speculate that a similar evolutionary process has taken
place for \source thus explaining the relatively short lifetime of the
NS through the dipole and wind spin-down and present RLOF
spin-up epochs.
 
A possible test for the AIC scenario can be performed by measuring the
gravitational mass of the NS in \source. Since the RLOF
phase has lasted for a very short time, the NS mass is
currently almost identical to the value expected from the AIC
scenario, which is $M\simeq1.25\msun$. A gravitational NS
mass larger than this value would immediately rule out the AIC scenario.

\acknowledgements{AP acknowledges support from the
  Netherlands Organization for Scientific Research (NWO) Veni
  Fellowship. MAA thanks the Astronomical Institut Anton Pannekoek for
  hospitality and the NWO for a grant during his sabbatical in
  Amsterdam. MAA is a member of the Science Academy, Istanbul, Turkey. }


\begin{thebibliography}{60}
\expandafter\ifx\csname natexlab\endcsname\relax\def\natexlab#1{#1}\fi
\expandafter\ifx\csname url\endcsname\relax
  \def\url#1{{\tt #1}}\fi
\expandafter\ifx\csname urlprefix\endcsname\relax\def\urlprefix{URL }\fi

\bibitem[{{Alpar}(2011)}]{alp11} 
{Alpar} M.A., 2011, \mnras, submitted

\bibitem[{{Alpar} \& {Psaltis}(2008)}]{alp08}
{Alpar} M.A., {Psaltis} D., Dec. 2008, \mnras, 391, 1472

\bibitem[{{Alpar} et~al.(1982){Alpar}, {Cheng}, {Ruderman}, \&
  {Shaham}}]{alp82}
{Alpar} M.A., {Cheng} A.F., et~al., Dec. 1982, \nat, 300, 728

\bibitem[{{Altamirano} et~al.(2008){Altamirano}, {Casella}, {Patruno},
  {Wijnands}, \& {van der Klis}}]{alt08}
{Altamirano} D., {Casella} P., et~al., Feb. 2008, \apjl, 674, L45

\bibitem[{{Altamirano} et~al.(2010){Altamirano}, {Homan}, {Linares}
  et~al.}]{alt10}
{Altamirano} D., {Homan} J., et~al., Oct. 2010, The Astronomer's Telegram,
  2952, 1

\bibitem[Bordas et al.(2010)]{bor10} Bordas, P., Kuulkers, 
E., Alfonso-Garz{\'o}n, J., et al.\ 2010, The Astronomer's Telegram, 2919, 
1 

\bibitem[{{Andersson} et~al.(2005){Andersson}, {Glampedakis}, {Haskell}, \&
  {Watts}}]{and05}
{Andersson} N., {Glampedakis} K., et~al., Aug. 2005, \mnras, 361, 1153

\bibitem[{{Bildsten} et~al.(1997){Bildsten}, {Chakrabarty}, {Chiu}
  et~al.}]{bilchak97}
{Bildsten} L., {Chakrabarty} D., et~al., Dec. 1997, \apjs, 113, 367

\bibitem[{{Boyles} et~al.(2011){Boyles}, {Lorimer}, {Turk} et~al.}]{boy11}
{Boyles} J., {Lorimer} D.R., et~al., Nov. 2011, \apj, 742, 51

\bibitem[{{Cavecchi} et~al.(2011){Cavecchi}, {Patruno}, {Haskell}
  et~al.}]{cav11}
{Cavecchi} Y., {Patruno} A., et~al., Oct. 2011, \apjl, 740, L8

\bibitem[{{Chenevez} et~al.(2010){Chenevez}, {Kuulkers}, {Alfonso-Garz{\'o}n}
  et~al.}]{che10}
{Chenevez} J., {Kuulkers} E., et~al., Oct. 2010, The Astronomer's Telegram,
  2924, 1

\bibitem[Cognrd et al.(1996)]{cog96} Cognard, I., Bourgois, G.,
  Lestrade, J.-F., et al. 1996, \aap, 311, 179

\bibitem[{{D'Angelo} \& {Spruit}(2010)}]{dan10}
{D'Angelo} C.R., {Spruit} H.C., Aug. 2010, \mnras, 406, 1208

\bibitem[{{Degenaar} \& {Wijnands}(2011)}]{deg11}
{Degenaar} N., {Wijnands} R., Jun. 2011, \mnras, 414, L50

\bibitem[{{Edgar}(2004)}]{edg04}
{Edgar} R., Sep. 2004, \nar, 48, 843

\bibitem[{{Eggleton}(1983)}]{egg83}
{Eggleton} P.P., May 1983, \apj, 268, 368

\bibitem[{{Ertan} et~al.(2009){Ertan}, {Ek{\c s}i}, {Erkut}, \&
  {Alpar}}]{ert09}
{Ertan} {\"U}., {Ek{\c s}i} K.Y., et~al., Sep. 2009, \apj, 702, 1309

\bibitem[{{Faucher-Gigu{\`e}re} \& {Kaspi}(2006)}]{fau06}
{Faucher-Gigu{\`e}re} C.A., {Kaspi} V.M., May 2006, \apj, 643, 332

\bibitem[{{Ferraro} et~al.(2009){Ferraro}, {Dalessandro}, {Mucciarelli}
  et~al.}]{fer09}
{Ferraro} F.R., {Dalessandro} E., et~al., Nov. 2009, \nat, 462, 483

\bibitem[{{Freire} et~al.(2011){Freire}, {Abdo}, {Ajello}, \& {et al.}}]{fre11}
{Freire} P.C.C., {Abdo} A.A., et~al., 2011, Science, 334, 1107

\bibitem[{{Ghosh} \& {Lamb}(1979)}]{gho79}
{Ghosh} P., {Lamb} F.K., Nov. 1979, \apj, 234, 296

\bibitem[{{Hartman} et~al.(2008){Hartman}, {Patruno}, {Chakrabarty}
  et~al.}]{har08}
{Hartman} J.M., {Patruno} A., et~al., Mar. 2008, \apj, 675, 1468

\bibitem[Hartman et al.(2009)]{har09} Hartman, J.~M., 
Patruno, A., Chakrabarty, D., et al.\ 2009, \apj, 702, 1673 

\bibitem[{{Hartman} et~al.(2011){Hartman}, {Galloway}, \&
  {Chakrabarty}}]{har11}
{Hartman} J.M., {Galloway} D.K., {Chakrabarty} D., Jan. 2011, \apj, 726, 26

\bibitem[{{Haskell} \& {Patruno}(2011)}]{has11}
{Haskell} B., {Patruno} A., Sep. 2011, \apjl, 738, L14

\bibitem[Heinke et al.(2006)]{hei06} Heinke, C.~O., Wijnands, 
R., Cohn, H.~N., et al.\ 2006, \apj, 651, 1098 

\bibitem[{{Hobbs} et~al.(2006){Hobbs}, {Edwards}, \& {Manchester}}]{hob06}
{Hobbs} G.B., {Edwards} R.T., {Manchester} R.N., Jun. 2006, \mnras, 369, 655

\bibitem[{{Hurley} et~al.(2002){Hurley}, {Tout}, \& {Pols}}]{hur02}
{Hurley} J.R., {Tout} C.A., {Pols} O.R., Feb. 2002, \mnras, 329, 897

\bibitem[Jahan-Miri(2000)]{jah00} Jahan-Miri, M.\ 2000, \apj, 
532, 514 

\bibitem[{{Jahoda} et~al.(2006){Jahoda}, {Markwardt}, {Radeva} et~al.}]{jah06}
{Jahoda} K., {Markwardt} C.B., et~al., Apr. 2006, \apjs, 163, 401

\bibitem[{{Kajava} et~al.(2011){Kajava}, {Ibragimov}, {Annala}, {Patruno}, \&
  {Poutanen}}]{kaj11}
{Kajava} J.J.E., {Ibragimov} A., et~al., Oct. 2011, \mnras, 417, 1454

\bibitem[{{Lamb} et~al.(1978){Lamb}, {Pines}, \& {Shaham}}]{lam78}
{Lamb} F.K., {Pines} D., {Shaham} J., Oct. 1978, \apj, 225, 582

\bibitem[Linares et al.(2012)]{lin12} Linares, M., 
Altamirano, D., Chakrabarty, D., Cumming, A., 
\& Keek, L.\ 2012, \apj, 748, 82 

\bibitem[{{Linares} et~al.(2011{\natexlab{b}}){Linares},
    {Chakrabarty}, \& {van der Klis}}]{lin11} {Linares} M.,
  {Chakrabarty} D., {van der Klis} M., Jun. 2011{\natexlab{b}}, \apjl,
  733, L17

\bibitem[Lyne et al.(1987)]{lyn87} Lyne, A.~g., Brinklow, A., Middleditch, J., Kulkarni, S.~R., \& Backer, D.~C.\ 1987, \nat, 328, 399


\bibitem[{{Lyne} et~al.(1996){Lyne}, {Manchester}, \& {D'Amico}}]{lyn96}
{Lyne} A.G., {Manchester} R.N., {D'Amico} N., Mar. 1996, \apjl, 460, L41

\bibitem[{{Markwardt} \& {Strohmayer}(2010)}]{str10}
{Markwardt} C.B., {Strohmayer} T.E., Jul. 2010, \apjl, 717, L149

\bibitem[{{Miller} et~al.(2011){Miller}, {Maitra}, {Cackett}, {Bhattacharyya},
  \& {Strohmayer}}]{mil11}
{Miller} J.M., {Maitra} D., et~al., Apr. 2011, \apjl, 731, L7

\bibitem[{{Miyaji} et~al.(1980){Miyaji}, {Nomoto}, {Yokoi}, \&
  {Sugimoto}}]{miy80}
{Miyaji} S., {Nomoto} K., et~al., 1980, \pasj, 32, 303

\bibitem[{{Nagae} et~al.(2004){Nagae}, {Oka}, {Matsuda} et~al.}]{nag04}
{Nagae} T., {Oka} K., et~al., May 2004, \aap, 419, 335

\bibitem[{{Nomoto}(1987)}]{nom87}
{Nomoto} K., Nov. 1987, \apj, 322, 206

\bibitem[{{Ortolani} et~al.(2007){Ortolani}, {Barbuy}, {Bica}, {Zoccali}, \&
  {Renzini}}]{ort07}
{Ortolani} S., {Barbuy} B., et~al., Aug. 2007, \aap, 470, 1043

\bibitem[{{Papitto} et~al.(2011){Papitto}, {D'A{\`i}}, {Motta} et~al.}]{pap11}
{Papitto} A., {D'A{\`i}} A., et~al., Feb. 2011, \aap, 526, L3

\bibitem[{{Patruno} et~al.(2010){Patruno}, {Hartman}, {Wijnands},
  {Chakrabarty}, \& {van der Klis}}]{pat10}
{Patruno} A., {Hartman} J.M., et~al., Jul. 2010, \apj, 717, 1253

\bibitem[Patruno et al.(2012)]{pat12} Patruno, A., Haskell, 
B., \& D'Angelo, C.\ 2012, \apj, 746, 9 

\bibitem[{{Pfahl} et~al.(2002){Pfahl}, {Rappaport}, \& {Podsiadlowski}}]{pfa02}
{Pfahl} E., {Rappaport} S., {Podsiadlowski} P., May 2002, \apjl, 571, L37

\bibitem[Podsiadlowski 
\& Mohamed(2007)]{pod07} Podsiadlowski, P., \& Mohamed, S.\ 2007, Baltic Astronomy, 16, 26 

\bibitem[{{Pooley} et~al.(2010){Pooley}, {Homan}, {Heinke} et~al.}]{poo10}
{Pooley} D., {Homan} J., et~al., Oct. 2010, The Astronomer's Telegram, 2974, 1

\bibitem[{{Psaltis} \& {Chakrabarty}(1999)}]{psa99}
{Psaltis} D., {Chakrabarty} D., Aug. 1999, \apj, 521, 332

\bibitem[{{Radhakrishnan} \& {Srinivasan}(1982)}]{rad82}
{Radhakrishnan} V., {Srinivasan} G., Dec. 1982, Current Science, 51, 1096

\bibitem[{{Rappaport} et~al.(2004){Rappaport}, {Fregeau}, \& {Spruit}}]{rap04}
{Rappaport} S.A., {Fregeau} J.M., {Spruit} H., May 2004, \apj, 606, 436

\bibitem[{{Reimers}(1975)}]{rei75}
{Reimers} D., 1975, Memoires of the Societe Royale des Sciences de Liege, 8,
  369

\bibitem[{{Reimers}(1978)}]{rei78}
{Reimers} D., 1978, Mitteilungen der Astronomischen Gesellschaft Hamburg, 43,
  70

\bibitem[{{Riggio} et~al.(2011){Riggio}, {Burderi}, {di Salvo} et~al.}]{rig11}
{Riggio} A., {Burderi} L., et~al., Jul. 2011, \aap, 531, A140

\bibitem[{{Romani}(1990)}]{rom90}
{Romani} R.W., Oct. 1990, \nat, 347, 741

\bibitem[{{Spitkovsky}(2006)}]{spi06}
{Spitkovsky} A., Sep. 2006, \apjl, 648, L51

\bibitem[{{Srinivasan} et~al.(1990){Srinivasan}, {Bhattacharya}, {Muslimov}, \&
  {Tsygan}}]{sri90}
{Srinivasan} G., {Bhattacharya} D., et~al., Jan. 1990, Current Science, 59, 31

\bibitem[{{Strohmayer} et~al.(2003){Strohmayer}, {Markwardt}, {Swank}, \& {in't
  Zand}}]{str03}
{Strohmayer} T.E., {Markwardt} C.B., et~al., Oct. 2003, \apjl, 596, L67

\bibitem[{{Theuns} \& {Jorissen}(1993)}]{the93}
{Theuns} T., {Jorissen} A., Dec. 1993, \mnras, 265, 946

\bibitem[{{Theuns} et~al.(1996){Theuns}, {Boffin}, \& {Jorissen}}]{the96}
{Theuns} T., {Boffin} H.M.J., {Jorissen} A., Jun. 1996, \mnras, 280, 1264

\bibitem[{{van den Heuvel}(2011)}]{van11}
{van den Heuvel} E.P.J., Mar. 2011, Bulletin of the Astronomical Society of
  India, 39, 1

\bibitem[{{van der Klis}(2001)}]{van01}
{van der Klis} M., Nov. 2001, \apj, 561, 943

\bibitem[{{Verbunt} et~al.(1990){Verbunt}, {Wijers}, \& {Burm}}]{ver90}
{Verbunt} F., {Wijers} R.A.M.J., {Burm} H.M.G., Aug. 1990, \aap, 234, 195

\bibitem[{{Watts} et~al.(2008){Watts}, {Patruno}, \& {van der Klis}}]{wat08}
{Watts} A.L., {Patruno} A., {van der Klis} M., Nov. 2008, \apjl, 688, L37

\bibitem[{{Wijnands} \& {van der Klis}(1998)}]{wij98}
{Wijnands} R., {van der Klis} M., Jul. 1998, \nat, 394, 344

\end{thebibliography}
\end{document}